\documentclass[fontsize=10pt, paper=letter, twocolumn]{scrartcl}
\usepackage[utf8]{inputenc}
\usepackage[english]{babel}
\usepackage{authblk}
\usepackage{siunitx}
\usepackage{amsmath}
\usepackage{amssymb}
\usepackage{graphicx}
\usepackage[includeheadfoot, headsep=.1in, margin={.6in,.6in}]{geometry}
\usepackage{tabularx}
\usepackage{hyperref}

\begin{document}

\title{Accurate evaluation of size and refractive index for spherical objects in quantitative phase imaging}

\author{Paul M\"{u}ller, Mirjam Sch\"{u}rmann, Salvatore Girardo, Gheorghe Cojoc, \\ and Jochen Guck\footnote{To whom correspondence should be addressed: jochen.guck@tu-dresden.de}}

\affil{Biotechnology Center of the TU Dresden, Germany}

\maketitle

This manuscript has been published in its final form at \url{https://dx.doi.org/10.1364/OE.26.010729}. The present pre-peer reviewed version of the manuscript contains an extended appendix which, due to a page-limit, did not find its way into the original publication.

\abstract{
Measuring the average refractive index (RI) of spherical objects, such as suspended cells, in quantitative phase imaging (QPI) requires a decoupling of RI and size from the QPI data.
This has been commonly achieved by determining the object's radius with geometrical approaches, neglecting light-scattering.
Here, we present a novel QPI fitting algorithm that reliably uncouples the RI using Mie theory and a semi-analytical, corrected Rytov approach.
We assess the range of validity of this algorithm \textit{in silico} and experimentally investigate various objects (oil and protein droplets, microgel beads, cells) and noise conditions.
In addition, we provide important practical cues for future studies in cell biology.
}

\section{Introduction}
Quantitative phase imaging (QPI) is a collective term for interferometric techniques that quantify the phase retardation of otherwise transparent objects.
The establishment of QPI as a swift tool in single-cell analysis has given rise to a broad spectrum of applications in cell biology.
For instance, QPI has been used to quantify cell dry mass \cite{Barer1952,davieswilkins1952}, cell dynamics \cite{Rappaz2005, Pavillon_2010, Shaked_2010, Jourdain_2011}, bacterial infection \cite{Ekpenyong_2012}, parasitic infection \cite{Park_2008}, or cellular differentiation \cite{Chalut2012}.
An important quantity that can be measured with QPI is the cellular refractive index (RI).
Besides characterizing cells based on RI and the associated local protein concentration \cite{Barer1952,davieswilkins1952,Barer1954a}, knowing the RI is also essential for related applications such as the optical stretcher to quantify optical forces \cite{Guck2001a, Boyde11, Boyde2012}, or for Brillouin microscopy to compute cell elasticity \cite{Scarcelli2015, Meng2016}.

The accurate determination of the cellular RI with QPI is difficult, because the optical path difference (OPD) measured in QPI needs to be separated into integral RI and cell thickness.
Thus, the integral RI can only be computed if the cell thickness is measured, which has been achieved using scanning probe microscopy \cite{Edward_2009, Zhang_2017}, confocal microscopy \cite{Curl_2005}, deliberate variation of the OPD \cite{Rappaz2005, Rappaz_2008}, or spatial confinement \cite{Lue_2006, Kemper2006}.
However, RI computation with such OPD approaches is based purely on geometric considerations, neglecting light-scattering.
A more general approach to this problem is optical diffraction tomography (ODT) \cite{Wolf1969,Sung2009}, which yields a complete 3D map of the intracellular RI. However, ODT depends on the acquisition of multiple phase images, which is accompanied by an elaborate experimental effort. Moreover, state-of-the-art ODT techniques approximate light propagation with the Rytov approximation \cite{Devaney1981, Slaney1984}, which can cause an underestimation of the intracellular RI for large RI gradients \cite{Mueller15, Schuermann2017}.

However, if a cell is spherical and sufficiently homogeneous in RI, then its average RI can be inferred from a single phase image.
Several studies have exploited cell sphericity by computing the cell radius from the cell area visible in the phase image using OPD approaches\cite{Kemper2007, Kemmler_2007, Schuermann2015, Steelman2017}.
However, a rigorous treatment of this problem by exactly modeling light propagation with Mie theory has not been presented so far due to the large computational effort required.

Here, we present a QPI phase fitting algorithm that reliably retrieves RI and size of spherical objects.
We address the computational challenge with efficient implementations of the Mie- and Rytov-scattered fields.
Additionally, we derive a correction factor for the Rytov approximation, resulting in accuracies similar to the results of Mie theory.
Using \textit{in silico} simulations, we compare our scattering approaches to OPD approaches for homogeneous spheres with RI values from 1.334 to 1.440.
In the experimental part, we demonstrate 2D QPI phase fitting for several test targets and for various noise conditions.
Our approach allows for a faster and better interpretation of QPI data for spherical objects and gives valuable insights into single-cell light-scattering.

\section{Methods}
\subsection{Modeling light-scattering by a sphere}
The OPD approach resembles a highly simplified model for light-scattering, treating the propagation of light as a line integral through the sphere's constant RI.
To estimate the average RI of a cell from a single quantitative phase image with the OPD approach, the cell radius  must be determined.
The radius can be determined with a series of 1D phase profile fits \cite{Kemper2007}, or by fitting a circle to a contour along the cell edge (found using e.g. the Canny edge detection algorithm) \cite{Schuermann2016}.
The former approach is equivalent to a direct fit of the projected RI of a sphere to the measured phase image (referred to as "OPD projection approach" in the remainder of the manuscript), while the latter approach only uses the contour data to determine the radius (referred to as "OPD edge-detection approach"). Note that the OPD edge-detection approach is somewhat erratic, because the contour found with an edge detection algorithm depends on image resolution.
Both OPD approaches do not take into account diffraction and thus only describe optically thin objects correctly.

The Rytov approximation gives a better estimation of light propagation \cite{Iwata1975}.
It is state-of-the-art in ODT applications, because it provides a simple linear model for diffraction and because it is sufficiently accurate for many cell types \cite{Mueller15}.
In general, to compute the Rytov approximation, a full 3D model of the specimen is required and its Fourier transform must be computed according to the Fourier diffraction theorem \cite{Wolf1969}.
In case of a sphere, however, we reduced this problem to two dimensions by using the analytical solution of the Fourier transform of a sphere (see appendix~\ref{ap:1} for a detailed description).
Our semi-analytical approach enables a fast and accurate computation of the Rytov approximation for homogeneous spheres.

Mie theory yields an exact solution for the scattered field. Here, we used the software BHFIELD\footnote{versioned October 5th 2012,  available at \url{https://seafile.zfn.uni-bremen.de/f/2d6fc70841/?dl=1} (accessed August 2017)} \cite{Suzuki2008} and refocused the obtained fields to the plane at the sphere center using the Python package nrefocus\footnote{version 0.1.5, available at \url{https://pypi.python.org/pypi/nrefocus}}. Generating the complete scattered field of a sphere using Mie theory is computationally expensive. Hence, we approximated the full 2D field by radially averaging two 1D fields, perpendicular and parallel to the polarization axis. As discussed in appendix~\ref{ap:2}, the error made by this approximation is negligible for the purpose of the present study.

\begin{figure}[ht]
\includegraphics[width=\linewidth]{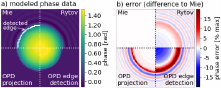}
\caption{Employed scattering models.}{\textbf{a)} Simulated quantitative phase images of a sphere with a radius of \SI{2.5}{\um} and  a refractive index (RI) of 1.36 embedded in a medium with an RI of 1.333 at a wavelength of \SI{550}{nm}. Each quadrant shows a phase image for one of the scattering models described in the text. The phase image of the optical path difference (OPD) edge-detection approach (lower right quadrant) was modeled with the OPD projection approach using the parameters obtained from the detected edge (edge shown in white in the upper left quadrant). \textbf{b)}~Difference between the scattering models and the Mie model (upper left quadrant in (a)) in \% of the maximum OPD.}
\label{fig:01}
\end{figure}

A comparison of these scattering models with Mie theory as a reference is shown in figure~\ref{fig:01} for an exemplary homogeneous sphere.
While the error made by the Rytov approximation is small, the error of the OPD projection approach scores high values, mostly due to the discontinuity in the gradient of the simulated phase image.
To allow a comparison to the OPD edge-detection approach as well, the lower right quadrants of figures~\ref{fig:01}a and~\ref{fig:01}b show the OPD projection and resulting error based on RI and radius as obtained with the OPD edge-detection approach applied to the Mie data. 
The OPD edge-detection approach exhibits a large error which can be explained by an underestimation of the radius due to the fact that the contour found by the edge detection algorithm does not coincide with the lateral perimeter of the sphere. This example illustrates that the determination of the RI and the radius of spherical objects should be addressed with models that take into account diffraction (Mie theory, Rytov approximation), rather than plain OPD approaches (OPD projection, OPD edge-detection).

\subsection{Fitting scattering models \textit{in-silico}}
The determination of the RI and the radius of a homogeneous sphere from a single quantitative phase image using one of the scattering models introduced above requires a phase image fitting algorithm. Previous work on this topic employed a set of 1D OPD fits to the 2D phase image \cite{Kemper2007}. For Mie theory or the Rytov approximation, this 1D approach is inefficient, either because fitting requires an enormous number of field computations (Mie) or because each iteration of the 1D fit requires the computation of a 2D field (Rytov, our implementation). To take advantage of the performance-enhancing procedures described in the previous section (Mie and Rytov), an actual 2D phase image fit is, in fact, necessary.

Here, we propose an image fitting algorithm that we specifically designed for homogeneous spheres in QPI.
Our algorithm iteratively fits RI, radius, and lateral position of a sphere as well as phase offset to a quantitative phase image.
In contrast to commonly used fitting techniques that call the modeling function for every parameter set during fitting, our algorithm interpolates the modeled phase image in a predefined interval and thus requires less calls to the modeling function.
This approach, which is described in detail in appendix~\ref{ap:3}, reliably retrieves RI, radius, lateral position, and phase offset for homogeneous, spherical phase objects\footnote{Our algorithm is part of the Python package qpsphere version 0.1.3, available at \url{https://pypi.python.org/pypi/qpsphere}.}.

\subsubsection{Error of the scattering models}
\label{sec:errormodel}
\begin{figure}[p]
\includegraphics[width=\hsize]{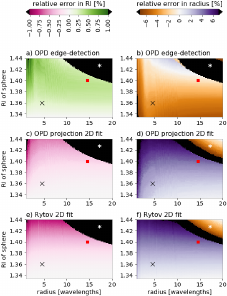}
\caption{Quantification of modeling errors.}{Left column: error for refractive index (RI) and right column: error for radius when using \textbf{(a,b)} the optical path difference (OPD) edge-detection approach or the proposed 2D fitting algorithm with either \textbf{(c,d)} the OPD projection approach or \textbf{(e,f)} the Rytov approximation. The ground truth data were generated from 9600 Mie simulations with an RI of the medium of 1.333 (water), and a grid size of 128$\times$128\,px. The lateral simulation size was $4r$ for $r<5\lambda$ (vacuum wavelength $\lambda$, sphere radius $r$) to capture diffraction effects and $3r$ for $r>10\lambda$ to enable Mie simulations for larger spheres, with a linear transition in-between. Initial values for the 2D fit were obtained using the OPD edge-detection approach. The cross and the square depict the spheres shown in figures~\ref{fig:01} and~\ref{fig:03}. The asterisk marks a region where 2D fitting fails due to large phase gradients (see text).}
\label{fig:02}
\end{figure}
To assess the accuracy of our 2D fitting approach, we performed a series of Mie simulations and retrieved RI and radius using the light-scattering models described above. Figure \ref{fig:02} shows the relative errors in RI and radius made using the OPD edge-detection approach and the 2D phase fitting approach (OPD projection, Rytov approximation).
The relative errors for RI ($n$) and radius ($r$) were computed using
\begin{align}
\Delta n &= \frac{n_\mathrm{measured} - n_\mathrm{exact}}{n_\mathrm{exact}},\\
\Delta r &= \frac{r_\mathrm{measured} - r_\mathrm{exact}}{r_\mathrm{exact}}.
\end{align}
The black cross and the red square label the positions of the exemplary spheres used in figure \ref{fig:01} and \ref{fig:03}. The white asterisk labels a region where the phase image resolution cannot resolve steep phase gradients ($>$2$\pi$ phase jump between two pixels) \cite{Steelman2017}. As a result, all approaches failed to determine the correct RI and radius in this region,  which we also observed with the otherwise error-free 2D Mie fit shown in figure~\ref{fig:S2} in the appendix.
As discussed above, the OPD edge-detection approach underestimates the radius of the sphere and thus overestimates the RI, which is clearly visible in figures~\ref{fig:02}a (green) and~\ref{fig:02}b (brown).
The 2D fit with the OPD projection (figs.~\ref{fig:02}c and~\ref{fig:02}d) exhibited a noisy error signal, which can be attributed to the discontinuity in the gradient of the modeled phase. In comparison, the error of the 2D Rytov fit shown in figures~\ref{fig:02}e and~\ref{fig:02}f is smooth and has lower values than the OPD projection in both RI and radius for RI values below 1.39.
The data suggest that the comparatively faster Rytov approximation can be preferred over Mie theory, if an error in the radius below 2\% and an error in the RI below 0.1\% is acceptable and if the imaged object has a radius above three wavelengths (3$\lambda$) and an RI below 1.36.

\subsubsection{Convergence with noise}
\label{sec:convnoise}
\begin{figure}[bh]
\includegraphics[width=\linewidth]{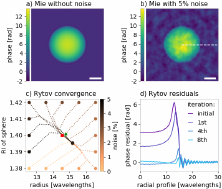}
\caption{Fitting robustness against noise.}{\textbf{a)} Phase image of a sphere with a radius of 14.5 wavelengths ($\lambda$) and a refractive index (RI) of $n_\text{sph}$\,=\,1.40 embedded in water ($n_\text{water}$\,=\,1.333) computed with Mie theory. Scale bar, 10\,$\lambda$. \textbf{b)}~Same phase image in (a) with \SI{5}{\%} noise added (see text for noise definition). \textbf{c)} For different initial parameters of RI and radius (dots at the border) and for several noise levels (color code), the phase images were fitted with our new fitting algorithm using the Rytov approximation. For each case, the fitting process, which took on average seven iterations, is shown as a dotted line. The red square indicates the exact value. The green triangle indicates the value obtained with the OPD edge-detection approach. \textbf{d)}~Radial residual profiles for the \SI{5}{\%} noise case (profile indicated in (b)) for selected iterations.}
\label{fig:03}
\end{figure}
In practice, quantitative phase images are subject to phase noise which can be described as a background pattern that is varying over the range of several pixels.
Here, we modeled phase noise using 2D Perlin noise as implemented in the Python package noise\footnote{version 1.2.1, available at \url{https://pypi.python.org/pypi/noise/}}.
Figure \ref{fig:03} illustrates the convergence of the proposed fitting algorithm for various starting parameters and for realistic noise conditions ranging from 0\% to 5\% standard deviation measured relative to the maximum OPD.
The data show that the algorithm converges to the same value after an average of seven iterations, independent of noise or initial conditions.
Unexpectedly, the RI and radius obtained with the OPD edge-detection approach, labeled with a green triangle, is closer to the correct value. This is a coincidental result that can be explained by the fact that the OPD edge-detection approach is resolution-dependent, i.e. the resolution chosen lead to better results for the OPD edge-detection approach (see fig.~\ref{fig:02}a,b).
Note that the initial conditions shown in figure~\ref{fig:03}c are rather extreme. In practice, we determined the initial guess for radius, RI, and position of the sphere with the OPD edge-detection approach. Thus, experimental phase noise does not affect the convergence of the proposed 2D phase image fitting algorithm.

\subsection{Systematic correction for the Rytov approximation}
\label{sec:syscorryt}
The errors in RI and radius made by the Rytov approximation shown in figures~\ref{fig:02}e and \ref{fig:02}f become large for objects with RI values above 1.36.
Thus, 2D fitting with the Rytov approximation would yield poor accuracies for many cell types with RI values reaching up to 1.40 and above.
In terms of efficiency, this would be a drawback, because high accuracies could only be achieved by falling back to the computationally more expensive Mie theory.
To address this issue, we derived a systematic correction for the Rytov approximation, making it possible to access large RI values with high accuracy.
We found that our implementation of the Rytov approximation (see appendix~\ref{ap:1}) exhibits a systematic error that is independent of the image resolution used. 
The major fraction of this systematic error is dependent only on the RI for object radii of three wavelengths and above.
This allowed us to derive a systematic correction for the Rytov approximation (Rytov-SC), considerably reducing the error made.
We derived the correction formulas for RI ($n_\text{Ryt-SC}$) and radius ($r_\text{Ryt-SC}$) by fitting a polynomial function to the Rytov error, yielding
\begin{align}
n_\text{Ryt-SC} &= n_\text{Ryt} + \left( 1.936 x^2 - 0.012 x \right) n_\text{med} \label{eq:rytscn}\\
r_\text{Ryt-SC} &=  r_\text{Ryt} \cdot  \left( -2.431 x^2 - 0.753x + 1.001 \right) \label{eq:rytscr}\\
&\text{with~} x = \frac{n_\text{Ryt}}{n_\text{med}} - 1
\end{align}
where $n_\text{med}$ is the RI of the medium and $n_\text{Ryt}$ is the RI obtained using our 2D fitting algorithm and our implementation of the Rytov approximation.
Note that we chose the variable $x$ such that the correction is independent of the RI of the medium $n_\text{med}$ (see figure~\ref{fig:S3} in the appendix). As we used the error maps shown in figure~\ref{fig:02} to derive equations~\ref{eq:rytscn} and~\ref{eq:rytscr}, the systematic correction is valid for spheres with RI $n_\text{sph}$ and radius $r_\text{sph}$ of at least
\begin{align}
1.0 &\leq \frac{n_\text{sph}}{n_\text{med}} \leq 1.08, \text{~and} \\
3\lambda &\leq r_\text{sph} \leq 20\lambda
\end{align}
with the vacuum wavelength~$\lambda$ of the light used.
This systematic correction extends the applicability of the Rytov approximation in 2D fitting to cells with high RI values (up to 1.44 and above), yielding high theoretical accuracies for RI ($<$0.1\%) and radius~($<$1\%).

\section{Results}
To compare the five approaches for the retrieval of RI and radius (OPD edge-detection, OPD projection fit, Rytov fit, Rytov-SC fit, Mie fit), we applied them to quantitative phase data of lipid droplets, microgel beads, and cells.

\subsection{2D Mie fits to doplets, beads, and cells}
\begin{figure}[p]
\includegraphics[width=\linewidth]{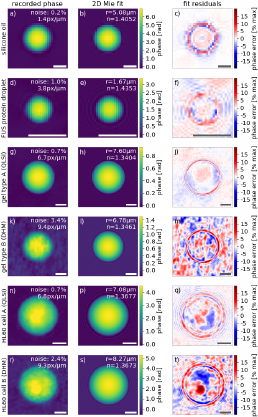}
\caption{2D Mie fits to experimental data.}{Experimental quantitative phase images (1st column), corresponding 2D fits with Mie theory (2nd column), and resulting fit residuals (3rd column) for liquid droplets (a-f), microgel beads (g-m) and HL60 cells (n-t). The scale bar is \SI{5}{\micro m}.}
\label{fig:04}
\end{figure}
Mie simulations in combination with our 2D fitting algorithm served as a benchmark for the other four approaches.
Figure~\ref{fig:04} shows a representative set of phase images, the corresponding Mie fits, and the fit residuals, which are discussed in the following.

Liquid droplets are ideal test samples for the investigation of scattering by spheres, because they are homogeneous and assume a spherical shape due to the surface tension at the droplet-medium interface.
Figure~\ref{fig:04}a shows a quantitative phase image of a silicone oil droplet embedded in phosphate buffered saline (PBS).
The oil droplet was produced by vortexing a two phase solution made of 1\,mL of silicon oil (Sigma-Aldrich, 10cSt) and 10\,mL of de-ionized water
containing 2\% w/v poly(ethylene glycol) monooleate (Sigma-Aldrich).
The image was recorded with quadriwave lateral shearing interferometry (QLSI) \cite{Bon_2009} using a commercial QPI camera (SID4Bio, Phasics S.A.) attached to an inverted microscope (AxioObserver Z1, Zeiss) with a 40$\times$ objective (NA 0.65, 421060-9900, Zeiss).
The illumination wavelength was confined to an average of 647\,nm using a bandpass filter (F37-647, 647/57, AHF analysentechnik).
We separately measured the RI of the silicone oil using an Abbe refractometer (2WAJ, Arcarda), yielding a value of 1.402 that matches the value of 1.405 from the 2D Mie fit shown in figure~\ref{fig:04}b.
The resulting relative error of 0.04\% is small which is reflected by residuals below 5\% of the maximum OPD shown in figure~\ref{fig:04}c.
Figure~\ref{fig:04}d shows the quantitative phase image of a protein droplet that was recorded with the same setup as above, except for the bandpass filter. Here we used an average imaging wavelength of 550\,nm for the RI analysis.
The droplet consisted of the RNA/DNA-binding protein FUS and was embedded in a protein buffer that is described in detail in reference \cite{Patel_2015}. FUS has been linked to neurodegenerative diseases such as amyotrophic lateral sclerosis (ALS) (see e.g. \cite{Polymenidou_2012, Wang_2013}) which is correlated to a liquid to solid phase transition of the protein with time \cite{Patel_2015}. Here, we obtained an average RI of 1.435 for a liquid FUS protein droplet (fig.~\ref{fig:04}e) using a protein buffer RI of  1.3465, measured with the Abbe refractometer named above. The fit residuals (fig~\ref{fig:04}f) are largely below 5\%, indicating good agreement of theory and experiment.
Note that the liquid droplets shown had comparatively high RI values and were imaged at low resolution, conditions that both  were handled well by the proposed 2D fitting algorithm.

Microgel beads offer an additional, convenient possibility to test scattering by a sphere. Here, we used polyacrylamide (PAA) beads that were produced in a flow focusing microfluidic system described elsewhere \cite{Christopher2007, Girardo2017}.
We imaged two types of PAA beads with different density (named A and B for simplicity) using two different imaging setups.
The microgel bead of type A was imaged with the QLSI setup used for the FUS droplet above with the addition of a telescope ($f_1$=$-15$\,mm, LD2020-A and $f_2$=75\,mm, LBF254-075-A, Thorlabs, Germany) to increase the magnification, and thus the sampling of the recorded phase image, by a factor of five.
The microgel bead of type B was imaged with digital holographic microscopy (DHM). A detailed description of the DHM setup used can be found in reference \cite{Schuermann2016}. A comparison of the resulting quantitative phase images in figures~\ref{fig:04}g and~\ref{fig:04}k shows that the DHM data have a higher background noise than the QLSI data (3.4\% vs. 0.7\%). The convergence of the 2D fitting algorithm is not affected by such noise magnitudes (see also fig.~\ref{fig:03}) as can be seen in the fit residuals that are either below 5\% (fig.~\ref{fig:04}j) or reproduce the background phase noise magnitudes (fig.~\ref{fig:04}m). 

HL60 cells in suspension have a spherical shape. Even though the internal RI distribution of a cell is generally non-uniform, a 2D Mie fit can be used to estimate its average RI. Figures~\ref{fig:04}n and~\ref{fig:04}r show quantitative phase images of two HL60 cells, recorded with the DHM and QLSI setups described above; To match the wavelength of the laser used for DHM imaging (HeNe 633\,nm, HNL050L-EC, Thorlabs), the illumination light of the QLSI setup was confined by a bandpass filter (BP 640/30, 488050-8001, Zeiss) in addition to filter F37-647 named above. Compared to liquid droplets or beads, the fit residuals shown in figures~\ref{fig:04}q and~\ref{fig:04}f exhibit heterogeneous structures that represent more and less dense regions within the cell. The similar RI values fitted to the representative cells, 1.3677 (QLSI) and 1.3673 (DHM), indicate consistency across the two different imaging setups and confirm resilience and accuracy of the 2D Mie fit with regard to phase noise, which is further discussed in the next section.

\subsection{Comparison of the scattering models}
\begin{figure}[bth]
\includegraphics[width=\linewidth]{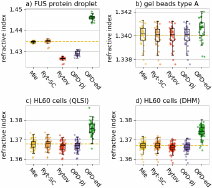}
\caption{Refractive index modeling results.}{Box plots of refractive index distributions determined for \textbf{a)} a time series of a FUS protein droplet (23 frames), \textbf{b)} microgel beads ($N$=39), \textbf{c)} HL60 cells measured with quadriwave lateral shearing interferometry (QLSI) ($N$=55), and \textbf{d)} HL60 cells measured with digital holographic microscopy (DHM) ($N$=84). The algorithms used are described in the text: Mie theory (yellow), the systematically corrected Rytov approximation using equation~\ref{eq:rytscn} (orange), the Rytov approximation (red), the OPD projection approach (blue) and the OPD edge-detection approach (green). Boxes extend from the lower to upper quartile values of the data, with a line at the median. Whiskers span 1.5$\times$ interquartile range. Dashed lines show the mean of the Mie data.}
\label{fig:05}
\end{figure}

Each of the methods presented to determine the average RI of spherical objects has benefits and drawbacks. Mie theory yields the best theoretically possible result, but it is computationally expensive. The Rytov approximation is less expensive and, with a systematic correction (equations~\ref{eq:rytscn} and~\ref{eq:rytscr}) it can achieve noteworthy accuracy. The OPD projection approach is faster than the Rytov approximation, but its accuracy is limited because it does not take into account diffraction. The same applies to the OPD edge-detection approach which is even faster but exhibits a resolution-dependent overestimation of the RI\footnote{In fact, a re-assessment of the average RI of a set of microgel beads presented in a previous publication shows that a 2D fit with the Rytov approximation ($n_\text{hydro2D-Ryt}$=1.354), when compared to the OPD edge-detection approach used in that publication ($n_\text{hydro2D-edge}$=1.356), is a better match for the microgel RI of an artificial cell phantom obtained using ODT ($n$=1.354) \cite{Schuermann2017}.}. To extend the theoretical insights obtained with the error maps shown in figure~\ref{fig:02}, a comparison of the scattering models based on four representative experimental data sets is shown in figure~\ref{fig:05}.

Figure \ref{fig:05}a shows the RI values determined for a single FUS protein droplet from a time series recorded with the setup introduced in the previous section. Note that, due to the low resolution of the phase images (fig.~\ref{fig:04}d), the RI values computed using the OPD edge-detection approach are severely overestimated and spread-out.

Figure~\ref{fig:05}b shows RI values for microgel beads of type A imaged with the corresponding QLSI setup described in the previous section (fig.~\ref{fig:04}g). Here, the difference between the scattering models is not as large as for the protein droplet, because the resolution is higher and because the average RI of the beads is much closer to the surrounding medium (PBS, $n_\text{PBS}$=1.335).

Figures \ref{fig:05}c and \ref{fig:05}d show the RI values fitted to two different HL60 cell populations recorded with QLSI and DHM (see figures~\ref{fig:04}n and~\ref{fig:04}r).
Note that each method produces consistent RI values across both imaging modalities (cell populations).
A comparison of the cell radii, shown in figure~\ref{fig:S4} in the appendix, indicates slight differences between the two populations, which is supposedly caused by biological variation.

In all examples shown, the systematically corrected Rytov approximation presents the best trade-off between accuracy and computational time.
The high accuracy becomes most evident for the case of the FUS protein droplet which had a large RI and thus caused a prominent systematic error in the Rytov approximation.
In case of the microgel beads, with RI values of about 1.34, the OPD projection approach and the non-corrected Rytov approximation still yield accurate results.
However, at RI values that are observed in HL60 cells, the systematic correction of the Rytov approximation becomes important, producing RI values that are considerably closer to Mie theory than all other approaches.
The total fitting time of the Rytov approximation is more than thirteen times shorter than the total fitting time of Mie simulations and less than three times longer than the fitting time of the OPD approach (e.g. for figure~\ref{fig:05}c: 17 minutes OPD projection, 48 minutes Rytov, 11 hours Mie) on a single CPU (Intel Core i7-4600U @2.10GHz, microcode version 0x21). 
Therefore, to accurately determine size and RI for spherical objects in QPI, we suggest the systematically corrected Rytov approximation in combination with a 2D fitting algorithm as presented here.

\section{Discussion}
\subsection{Summary}
\begin{figure}[htb]
\includegraphics[width=\linewidth]{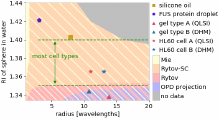}
\caption{Accuracy of the scattering models.}{The shaded regions indicate the range of validity for each model with a relative error in refractive index (RI) $\Delta n$ below 0.1\% and a relative error in radius $\Delta r$ below 1\% (see eqns.~\ref{eq:rytscn} and~\ref{eq:rytscr}). The OPD edge-detection approach is excluded because of its strong dependence on image resolution. The range of average RI values found for most cells is indicated in green (see table~\ref{tab:S1} in the appendix, literature examples). 
To provide orientation to the reader, the RI values (normalized to water by multiplication of $n_\text{sph}$ with $n_\text{water}/n_\text{med}$ with $n_\text{water}$=1.333) and radii (converted to wavelengths) of the samples shown in figure~\ref{fig:04}, are displayed as individual labels.
}
\label{fig:06}
\end{figure}

The present study provides an efficient method to accurately compute the RI and the radius for spherical objects from a single quantitative phase image.
We investigated OPD approaches (OPD edge-detection, OPD projection), which were used in previous publications, and introduced efficient implementations of diffraction-based models (Mie theory, systematically corrected Rytov approximation) in combination with a 2D phase fitting algorithm, yielding improved accuracy and stability.
Figure~\ref{fig:06} illustrates the validity ranges of the models used (shaded regions) and provides orientation by indicating the exemplary samples shown in figure~\ref{fig:04}.
We identified the systematically corrected Rytov approximation (Rytov-SC) as a feasible model for everyday-use in basic research, offering high accuracy at low computational costs over a broad range of sample sizes and RI values. Our 2D phase fitting approach can be applied to most biological cells (taking cell sphericity as given), is resilient to phase noise (see fig.~\ref{fig:03}), and operates well even at low image resolution (see fig.~\ref{fig:04}a-f).

\subsection{Outlook}
The presented combination of the 2D fitting algorithm and the systematically corrected Rytov approximation for a sphere is efficient, but still offers room for improvement. First, graphical processing units (GPUs) could significantly speed-up the Fourier transform and the image interpolation steps of our current implementation. Second, the choice for the sampling of the radius with 42 points to compute the Rytov approximation, as discussed in appendix~\ref{ap:1}, is conservative and could be reduced to about 20 points. In this case, however, the correction formulas (eqns.~\ref{eq:rytscn} and~\ref{eq:rytscr}) need to be updated. With these modifications, the fitting algorithm could in principle achieve real-time performance for live QPI analysis.

The fit residuals shown in figure~\ref{fig:04} exhibit small systematic phase errors, visible as blue and red circles around the object perimeter.
These can be attributed to the fact that the point spread function (PSF) of the imaging setups used is not taken into account in the simulation. At the cost of more computation time, the simulation images could in principle be convolved with the PSF which would yield lower residuals. However, we do not expect higher fitting accuracies from this approach, because the residuals shown are already low and because the number of pixels in the described region is small compared the total number of pixels that resemble the imaged object.

To resolve large intracellular compartments or to analyze cells that have an approximately ellipsoidal shape, our algorithm could be extended to support a superposition of spheres \cite{Jin2014} and ellipsoids. This would allow to resolve subcellular compartments such as the nucleus and the nucleolus from a single phase image or enable a frame-by-frame analysis of cell volume and RI of elongated cells in an optical stretcher experiment. Such extensions would yield an enhanced picture of the imaged cell at the cost of additional fitting parameters.

An accurate measurement of the RI of spherical objects is an important prerequisite for emerging topics in adjacent fields of research.
For instance, our approach could allow valuable insights into the aging process of FUS \cite{Patel_2015} by tracking and characterizing the fit residuals during the liquid to solid phase transition of individual protein droplets.
This would allow to determine whether the aging process starts at the center of the drop, at its surface, or homogeneously throughout.
Furthermore, our algorithms could be used to assess the quantitative imaging quality of ODT and QPI techniques, both of which frequently employ microgel or polymer beads as reference samples (e.g. \cite{Bon_2009, Sung2009, Schuermann2016, Steelman2017, Schuermann2017}). Finally, complementing the optical analysis of microgel beads as presented here with a mechanical description using, for instance, atomic force microscopy, will allow to establish a well-characterized microgel bead toolbox~\cite{Girardo2017} which would be an invaluable reference for the optomechanical analysis of cells using techniques such as Brillouin microscopy or the optical stretcher. Thus, the optical characterization of spheres is fundamentally important to address topical questions in cell biology and biophysics.

\section{Conclusion}
The methods presented here resemble an important foundation for the accurate characterization of spherical micro-objects, including the optomechanical phenotyping of live cells, the study of protein droplets, or the classification of artificial beads.
The presented approach is complementary to tomographic techniques, limited to spherical objects but able to deliver average values for RI and radius using only a single quantitative phase image.
The derivation of an efficient model for light scattering by a sphere and the implementation of an automated 2D phase-fitting algorithm resemble a major advancement in accuracy, stability, and throughput.
Hence, the presented approach is an attractive tool for many emerging techniques that rely on an exact optical characterization of spherical objects to address biophysical questions in basic and applied research.

\section*{Funding}
This project has received funding from the European Union’s Seventh Framework Programme for the Starting Grant "Light Touch" (grant agreement no. 282060) and from the Alexander-von-Humboldt Stiftung (Humboldt-Professorship to J.G.).

\section*{Acknowledgments}
The authors want to thank Tony Hyman, Simon Alberti, and their respective research groups for discussions as well as the Protein Expression and Purification Facility of the Max Planck Institute of Molecular Cell Biology and Genetics (MPI-CBG) for the provision of the FUS protein sample.
We thank the BIOTEC/CRTD Microstructure Facility (partly funded by the State of Saxony and the European Fund for Regional Development - EFRE) for the production of the microgel beads.
The HL60/S4 cells were a generous gift from Donald and Ada Olins (University of New England).
\appendix
\section{Appendix}
The algorithms presented in \ref{ap:1} to \ref{ap:4} are implemented in the Python package qpsphere version 0.1.3, available at \url{https://pypi.python.org/pypi/qpsphere}.

\subsection{The Rytov near-field of a sphere}
\label{ap:1}
We obtained the field scattered by a sphere in the Rytov approximation by computing its analytical 2D Fourier transform followed by an inverse Fourier transform using the Fourier diffraction theorem. The use of an analytical solution in Fourier space has the advantage that it reduces artifacts that arise from the sharp boundaries of the spherical volume in real space and that the 3D problem is broken down to a 2D problem.
According to the Fourier diffraction theorem, the Fourier transform of the 2D background-corrected scattered field $\widehat{U}_\mathrm{B}(\mathbf{k})$ is mapped onto a semi-spherical surface in the Fourier transform of the 3D object potential $\widehat{F}(\mathbf{k})$ according to \cite{Wolf1969, Kak2001, Mueller15arxiv}
\begin{equation}
\widehat{U}_\mathrm{B}(\mathbf{k_D}) = 
\frac{i \pi }{\sqrt{2\pi}}
\frac{
 \exp(i k_\mathrm{m} l_\mathrm{D}(M-1))
 }{
 k_\mathrm{m}M
 }    						
\widehat{F}\left(\mathbf{k_D} -k_\mathrm{m} \mathbf{s_0}\right),
\label{eq:FourierDiffract}
\end{equation}
where we used the notation of \cite{Mueller15arxiv} (unitary angular frequency Fourier transform) with the Fourier transformed detector coordinates $\mathbf{k_D}$, the distance between the center of the object potential and the detector plane $l_\mathrm{D}$, the wave number $k_\mathrm{m} = 2\pi n_\mathrm{med} / \lambda$ (refractive index of medium $n_\mathrm{med}$ and vacuum wavelength $\lambda$), the factor $M=\sqrt{1-k_\mathrm{x}^2-k_\mathrm{y}^2}/k_\mathrm{m}$, and the unit vector representing the direction of illumination, i.e. the rotational position of the sample relative to the detector normal, $\mathbf{s_0}$. 
The rotationally symmetric Fourier transform of a homogeneous sphere with radius $R$ is given by
\begin{equation}
\widehat{F}_\mathrm{sph}(k) = -\frac{2R^2}{\sqrt{2\pi}k} \cdot j_1(kR)
\label{eq:FourierSphere}
\end{equation}
where $j_1$ is the spherical Bessel function of the first kind of order one. Evaluating equation \ref{eq:FourierSphere} at the frequencies that correspond to the spherical surface described by equation \ref{eq:FourierDiffract} and performing an inverse Fourier transform of the resulting $\widehat{U}_\mathrm{B,sph}$ yields the background-corrected scattered field component in the Born approximation $u_\mathrm{B,sph}$. The Rytov approximation of the scattered field component $u_\mathrm{R,sph}$ is then obtained by computing the exponential of $u_\mathrm{B,sph}$ \cite{Kak2001, Chen:98}
\begin{equation}
u_\mathrm{R,sph}(\mathbf{r}) = \exp \left(\mathcal{F}^{-1} \left\lbrace\widehat{U}_\mathrm{B,sph}(\mathbf{k_D}) \right\rbrace \right)
\label{eq:RytovScatter}
\end{equation}
with the inverse Fourier transform operator $\mathcal{F}^{-1}$. Assuming a incident plane wave, the full complex field then computes to
\begin{equation}
u_\mathrm{sph}(\mathbf{r}) = 1 + u_\mathrm{R,sph}(\mathbf{r}).
\end{equation}
In practice, we computed the Rytov approximation in two steps. First, the scattered field is computed with a resolution that samples the sphere radius with approximately 42 points. Second, the amplitude and phase data are linearly interpolated to match the resolution of the images recorded in the experiment. This two-step approach ensures that the field computation is resolution-independent, allowing to derive a systematic error correction for the Rytov approximation to yield results comparable to Mie theory as presented and discussed in section~\ref{sec:syscorryt}.

\subsection{Efficient Mie computation by averaging}
\label{ap:2}
\begin{figure}[ht]
\includegraphics[width=\linewidth]{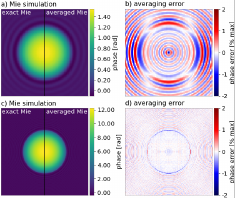}
\caption{Mie theory: exact vs. averaging.}{\textbf{a)} The left part of the image shows the exact Mie simulation and the right part the averaged Mie simulation for the dielectric sphere used in figure~\ref{fig:01}. \textbf{b)} The illustration of the averaging error shows a variation between $\pm 1\%$ relative to the maximum optical path difference. \textbf{c)} Comparison of the exact and averaged Mie simulation for figure~\ref{fig:03}. \textbf{d)} Illustration of the corresponding averaging error.}
\label{fig:S1}
\end{figure}

To compute the field scattered by a sphere, an exact solution based on Mie theory can be used \cite{Suzuki2008, Ringler2008}. However, compared to linear models, e.g. the optical path difference (OPD) projection or the Rytov approximation, Mie theory is computationally expensive. Here, we propose an averaged Mie approach, where the cross-sectional fields along a line perpendicular and parallel to the polarization of the simulated light are averaged. This approach reduces the required Mie field computation from a 2D image to two 1D lines, resulting in an effectively unpolarized field. Figure \ref{fig:S1} illustrates the error made using this approach, which is oscillating within the $\pm 1\%$ interval relative the the maximum optical path difference and can thus be neglected for the studies presented here.

\subsection{A 2D fitting algorithm for spheres in QPI}
\label{ap:3}
The 2D fitting of a phase image with a sphere-scattering model is burdened by many calls of the modeling function. In practice, this bottleneck becomes problematic when the scattering model is computationally expensive which is the case for e.g. Mie simulations. Here, we used a custom fitting algorithm that relies on interval-based phase image interpolation to reduce the number of calls to the modeling function. The algorithm reliably fits Mie, Rytov, and OPD projection models to a 2D phase image of a sphere with the five parameters
\begin{itemize}
\item[$\phi_\mathrm{bg}$] background phase image offset,
\item[$x$] x-coordinate of the center of the sphere,
\item[$y$] y-coordinate of the center of the sphere,
\item[$n_\mathrm{sph}$] refractive index (RI) of the sphere, and
\item[$r_\mathrm{sph}$] radius of the sphere.
\end{itemize}
Given initial estimates for center ($x_0$, $y_0$), RI $n_0$, and radius $r_0$ of the sphere, the algorithm iteratively fits a sphere model in six steps.
\begin{enumerate}
\item \textbf{Vary radius.} First, compute three phase images using the sphere model at the radii, $r_i-b_{r}$, $r_i$, and $r_i+b_r$, where initially $i=0$ and $b_r=0.05r_0$. Second, generate 47 phase images in the interval $[r_i-b_{r}, r_i+b_{r}]$ by interpolation. Third, compare the interpolated phase images with the measured phase image using the root mean square (RMS) error. The radius with the lowest RMS error $r_{i+1}$ is used for the following steps.
\item \textbf{Vary RI.} As in step one, compute three phase images with the sphere model at the RI values, $n_i-b_{n}$, $n_i$, and $n_i+b_n$, where initially $b_n=0.1(n_0-n_\mathrm{med})$ (RI of the medium $n_\mathrm{med}$). Then, interpolate 47 phase images in the interval and select the one with the lowest RMS error $n_{i+1}$.
\item \textbf{Vary center.} Initially, the interval parameter for the center coordinate $b_\mathrm{c}$ is set to one wavelength ($\lambda$) or 5\% of the radius, depending on which is larger
\begin{equation}
b_\mathrm{c} = \max ( \lambda, \,0.05r_0 ).
\end{equation}
Then, 13$\times$13 phase images are computed for the intervals $[x_i-b_\mathrm{c}, x_i+b_\mathrm{c}]$ and $[y_i-b_\mathrm{c}, y_i+b_\mathrm{c}]$. The phase image with the lowest RMS defines the center position $(x_{i+1}, y_{i+1})$ for subsequent iterations.
\item \textbf{Phase background estimation.} Experimental phase images can contain a constant phase offset. We estimate the phase offset $\phi_i$ by averaging the background phase values of the experimental phase image. To determine the pixel locations of the background phase image, the intersection of two pixel sets (a) and (b) is used:
\begin{enumerate}
\item[a)] all pixels within a frame border of the phase image with a width of either 5 pixels or a fifth of the image size, depending on which is larger, and
\item[b)] all pixels of the simulated phase whose absolute value is lower than 1\% of the maximum simulated phase.
\end{enumerate}
\item \textbf{Scale down interval parameters.} The interval parameters $b_r$, $b_n$, and $b_\mathrm{c}$ are individually divided by two, depending on the following conditions:
\begin{itemize}
\item[$b_r$:] The radius $r_{i+1}$ is in the interval ${[r_i-0.1b_{r}, r_i+0.1b_{r}]}$.
\item[$b_n$:] The RI $n_{i+1}$ is in the interval ${[n_i-0.1b_{n}, n_i+0.1b_{n}]}$.
\item[$b_\mathrm{c}$:] The change in the location of the center position $\sqrt{(x_i-x_{i+1})^2 + (y_i-y_{i+1})^2}$ is smaller than $b_\mathrm{c}$.
\end{itemize}
Scaling down the interval parameters leads to a parameter refinement in each iteration of the algorithm.
\item \textbf{Stopping criteria.} The algorithm stops iterating when all of the following conditions are met:
\begin{itemize}
\item The change in radius $|r_{i+1}-r_i|/r_{i+1}$ is smaller than 0.1\%.
\item The change in RI $|n_{i+1}-n_i|$ is smaller than 0.0005.
\end{itemize}
If the stopping conditions are not all met, then the algorithm proceeds with step one.
\end{enumerate}
As discussed in section~\ref{sec:convnoise}, the initial fitting parameters can be obtained with the OPD edge-detection approach. 

\begin{figure}[h]
\includegraphics[width=\linewidth]{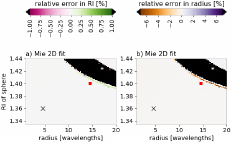}
\caption{Resolution limit of 2D fitting.}{Errors for refractive index and radius compared to the Mie simulation when using a Mie model in the proposed 2D phase fitting algorithm. For details see figure~\ref{fig:02}.}
\label{fig:S2}
\end{figure}
To demonstrate the correct convergence of the fitting algorithm, figure \ref{fig:S2} shows the 2D Mie fits to the Mie simulations, as is done in figure~\ref{fig:03} for OPD edge-detection, OPD projection, and Rytov approximation. Besides the fitting errors due to resolution-impaired unwrapping problems (region labeled with white asterisk), which is discussed in section~\ref{sec:errormodel}, the Mie fitting error is negligible.

\subsection{The corrected Rytov approximation}
\label{ap:4}
As discussed in section~\ref{sec:syscorryt}, the Rytov approximation, implemented as described above, exhibits a resolution-independent systematic error compared to Mie theory. Thus, we could derive a correction factor (equations~\ref{eq:rytscn} and~\ref{eq:rytscr}) for the Rytov approximation, yielding fitting accuracies close to Mie theory with comparatively little computational effort.

\begin{figure}[ht]
\includegraphics[width=\linewidth]{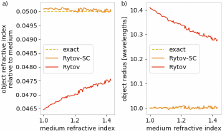}
\caption{Validity of corrected Ryotv approximation.}{The systematic correction for the  Rytov approximation given by equations~\ref{eq:rytscn} and~\ref{eq:rytscr} is independent of the refractive index of the medium as shown for \textbf{a)} the refractive index of the object and \textbf{b)} the radius of the object. The exact data to which the Rytov approximation was fitted were computed using Mie theory for a sphere with a radius ten wavelengths and a resolution of five pixels per wavelength on a grid of 150$\times$150 pixels.}
\label{fig:S3}
\end{figure}

To ensure that the systematic correction for the Rytov approximation is independent of the RI of the medium, we fitted the Rytov approximation to a set of Mie simulations with varying RI of the medium. Figure~\ref{fig:S3} shows the fitting results of the Rytov approximation and its systematic correction for a sphere with a radius of ten wavelengths and a constant RI difference between the sphere and the medium of 0.05.

\begin{figure}[ht]
\includegraphics[width=\linewidth]{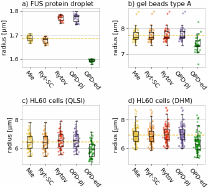}
\caption{Radius modeling results.}{Box plots of object radii distributions determined for \textbf{a)} a time series of a FUS protein droplet, \textbf{b)} microgel beads 	, \textbf{c)} HL60 cells measured with a quadriwave lateral shearing interferometry (QLSI), and \textbf{d)} HL60 cells measured with digital holographic microscopy (DHM). For more information see figure~\ref{fig:05}. Boxes extend from the lower to upper quartile values of the data, with a line at the median. Whiskers span 1.5$\times$ interquartile range. Dashed lines show the mean of the Mie data.}
\label{fig:S4}
\end{figure}

As for the fitted RI shown in figure~\ref{fig:05}, figure \ref{fig:S4} shows the fitted radii for all algorithms presented in our study.
The fit with the systematically corrected Rytov approximation (Rytov-SC) consistently yields radii resembling those of the Mie theory fit when compared to the non-corrected Rytov approximation fit.

\clearpage
\subsection{Cell refractive index values in literature}
\label{ap:5}
Cells can exhibit a wide range of RI values. Table \ref{tab:S1} lists the average RI of several cell types with literature references. Note that the differences in the RI values between subsequent publications (e.g. pancreas carcinoma cells: 1.375 versus 1.380) can be explained by large standard deviations (data not shown here). Overall, most cell types have an RI value that resides in the interval above 1.35 and below 1.40, which we used to indicate the biological relevance of this study in figure~\ref{fig:06}.
\begin{table}[ht]
\begin{tabularx}{\linewidth}{|l|X|}
\hline 
RI value & cell type\\ 
\hline 
\hline 
1.3568 & lung adenocarcinoma cells (A549) \cite{Steelman2017} \\ 
\hline
1.362 & myeloid leukemia cells (HL60/S4), differentiation towards neutrophils \cite{Chalut2012} \\ 
\hline 
1.366 & myeloid leukemia cells (HL60/S4) \cite{Chalut2012} \\ 
\hline
1.3662 & transformed lung epithelial cells (BEAS-2B) \cite{Steelman2017} \\ 
\hline
1.3671 & lymphoblastic leukemia cells (Jurkat) \cite{Schuermann2016} \\ 
\hline
1.369 &  liver tumor cell (HepG2) \cite{Kemper2007} \\
\hline
1.3713 &  invasive ductal carcinoma cells (MCF-7) \cite{Schuermann2016} \\ 
\hline
1.3716 &  cervical adenocarcinoma cells (HeLa) \cite{Schuermann2016} \\ 
\hline
1.3730 &  vaginal epithelium cells (HVE) \cite{Steelman2017} \\ 
\hline
1.375 &  pancreas carcinoma cells (Patu 8988S and Patu 8988T) \cite{Kemper2007, Kemper2011}; mouse cortical neurons \cite{Rappaz2005} \\ 
\hline
1.3775 &  invasive ductal carcinoma cells (MCF-7) \cite{Steelman2017} \\ 
\hline
1.3776 &  myeloid leukemia cells (HL60/S4) \cite{Schuermann2016} \\ 
\hline 
1.380 &  pancreas carcinoma cells (Patu 8988) \cite{Kemper2006} \\ 
\hline 
1.385 & mouse cortical neurons \cite{Rappaz2005} \\ 
\hline
1.399 &  red blood cells \cite{Park_2008}\\
\hline
\end{tabularx}
\caption{Average refractive index of several cell types.}
\label{tab:S1}
\end{table}

\bibliographystyle{ieeetr}
\bibliography{article}

\end{document}